\documentclass{elsarticle}
\usepackage{natbib}
\usepackage{enumerate}
\usepackage{amsmath,amsthm}
\usepackage{amssymb, latexsym}
\usepackage[mathscr]{euscript}
\usepackage[toc, page]{appendix}
\usepackage{hyperref}
\usepackage{color}
\usepackage{array}
\usepackage{flafter}
\usepackage{layout}
\usepackage{graphicx}
\usepackage{epstopdf}
\usepackage[ansinew]{inputenc}
\usepackage{xfrac}
\usepackage{algorithm}
\usepackage{soul}
\theoremstyle{plain}

\theoremstyle{definition}

\theoremstyle{remark}
  \newtheorem{remark}{Remark}
  
  \makeatletter
\def\ps@pprintTitle{%
   \let\@oddhead\@empty
   \let\@evenhead\@empty
   \def\@oddfoot{\reset@font\hfil\thepage\hfil}
   \let\@evenfoot\@oddfoot
}
\makeatother
 
 \newcommand*{\defeq}{\mathrel{\vcenter{\baselineskip0.5ex \lineskiplimit0pt
                     \hbox{\scriptsize.}\hbox{\scriptsize.}}}%
                     =}

\begin{document}

\author[1]{Daniel Andr\'es D\'{\i}az--Pach\'on} \ead{Ddiaz3@miami.edu}
\author[1]{J. Sunil Rao\corref{cor1}} \ead{JRao@biostat.med.miami.edu}
\cortext[cor1]{Corresponding author}
\address[1]{Division of Biostatistics - University of Miami, Don Soffer Clinical Research Center, 1120 NW 14th St, Miami FL, 33136}

\begin{abstract}
COVID-19 testing has become a standard approach for estimating prevalence which then assist in public health decision making to contain and mitigate the spread of the disease.  The sampling designs used are often biased in that they do not reflect the true underlying populations. For instance, individuals with strong symptoms are more likely to be tested than those with no symptoms. This results in biased estimates of prevalence (too high). Typical post-sampling corrections are not always possible.  Here we present a simple bias correction methodology derived and adapted from a correction for publication bias in meta analysis studies. The methodology is general enough to allow a wide variety of customization making it more useful in practice. Implementation is easily done using already collected information. Via a simulation and two real datasets, we show that the bias corrections can provide dramatic reductions in estimation error.  

\end{abstract}
\title{A simple correction for COVID-19 sampling bias}
\maketitle

\section{Introduction}

There is an urgent need to better understand the spread of COVID-19 in populations both from being able to identify important changes in infection dynamics but also in understanding the effectiveness of control and mitigation strategies. So testing studies of multiple sorts have been undertaken all over the world. Serological surveys (sometimes together with nucleic acid amplification testing) have become a widespread tool to estimate SARS-CoV-2 prevalence and to assess the extent the infection has spread in the population. On the other hand, the results of nucleic acid amplification testing are usually used for diagnosis or detection of SARS-CoV-2 infection and estimating the incidence of infections. Outside of a few exceptions however, most of these studies have been with biased samples. These can be convenience samples which can lead to over representation of symptomatic sampled units  \cite{AllevaEtAl2020}. And this in turn can lead to over-estimation of disease prevalence. Random sampling protocols can be inefficient due to the lower degree of infection amongst asymptomatic individuals. Thus there is a great need effective corrections that can reduce bias in prevalence estimation.

Much attention has been focused on the issue of correcting for imperfect tests \cite{Diggle2011, Greenland1996}; but less attention has been paid to correcting for biased sampling.   One notable exception is \cite{AllevaEtAl2020}, where it was proposed a snowball sampling approach in conjunction with contact tracing in order to set up a better disease surveillance system.  However, this is not how the vast majority of studies are conducted today.  In this paper, we address biased sampling from an entirely and somewhat unexpected viewpoint.  

We note that biased samples also occur when doing meta analyses due to publication bias \cite{AndrewsKasy2019}.  That is, papers which favor a null hypothesis of no treatment effect are less likely to be published and hence meta-analytic estimates of treatment effect can be over-estimated.  If the null-favoring censoring mechanism can be modeled, then interesting corrections can be made.  Based on \cite{AndrewsKasy2019}, we derive and adapt a version of their model for correcting sampling bias in the current COVID-19 pandemic. 

We develop the main idea taking into account several categories of symptoms. However, some readers might find of interest to consider just two categories (symptomatic and asymptomatic). In this case, a simple summary of the theory can be found in Subsection \ref{M2} and a real-life example for this situation can be found in Section \ref{cruise}.

\section{The model}

Consider a population $P$ of size $N$. $P$ has a partition $\mathcal P$ into $2M$ subsets of $P$ each with proportions given by the vector $\textbf p^* \defeq \left(p_1^{(0)}, p_1^{(1)}, p_2^{(0)}, p_2^{(1)}, \ldots, p_M^{(0)}, p_M^{(1)} \right)$, where

\begin{align*}
 	\sum_{s=1}^M \left( p_s^{(0)} + p_s^{(1)} \right) = 1 \text{ and } p_s^{(0)}, p_s^{(1)} \geq 0\text{, for } s \in \{1, \ldots, M\}. 
 \end{align*}
In other words, $\textbf p^*$ belongs to the standard $(2M-1)$-simplex. We define a r.v.\ $S^*$ taking values in the set $\textbf I \defeq \left\{1^{(0)}, 1^{(1)}, \ldots, M^{(0)}, M^{(1)}\right\}$ that, conditioned on $\textbf p^*$, selects an element of the partition $\mathcal P$ according to a categorical distribution in the interval $(0,1)$:
\begin{align*}
	f_{S^*} \left(s^{(i)} | \textbf p^* \right) = p_s^{(i)},
\end{align*}
where $s \in \{1, \ldots, M\}$ represents \textit{symptomatology} (i.e., number or degree of symptoms), and $i \in\{0,1\}$, \textit{prevalence}: $i=1$ represents infected, while $i=0$ represents non infected. Think of $\textbf I$ as the set of indexes of the elements in the partition $\mathcal P$. Thus, with a slight abuse of notation, we will refer to $s$ either as a category or as the subset of $\mathcal P$ with that particular index. In the most common scenario there are just the categories ---asymptomatic and symptomatic---, so $M=2$. However, some studies have considered more than two degrees of symptoms (see \cite{SudreEtAl2020}).

The proportion of people with symptoms $s = s^{(0)} \cup s^{(1)}$ is given by 
\begin{align}\label{symptomS}
p_s = p_s^{(0)} + p_s^{(1)}.
\end{align}
Here $p_s^{(1)}$ represents the probability of being in the category $s$ \underline{and} being infected, whereas $p_s^{(0)}$ represents the probability of being in the category $s$ while non infected.

From this notation, the overall probability of being infected is 
\begin{align*}
p_1^{(1)} +  \cdots + p_M^{(1)}.
\end{align*} 
Also, $p_s^{(1)}/p_s$ is the conditional probability of being infected given that we are considering the category $s$ of symptoms. 

We assume a Bernoulli r.v.\ $T$, which will be 1 with probability $p(S^*)$. Let's consider an independent sequence $T_1,\ldots, T_N$, distributed as $T$. If the individual $j$ belongs to the group $s^*$, $T_j=1$, which happens with probability $p(s^*)$, will tell us that the individual $j$ is tested (sampled).  The sample size is given by $N_T = \sum_{j=1}^N T_j$.  

In summary, up to this point, for $m \in \{1\ldots, M\}$, we have that:
\begin{itemize}
	\item $p(m)$ is the probability of being in the category $m$ and being tested.
	\item $p \left(m^{(1)} \right)$ is the probability of testing for an individual in category $m$  who is infected.
	\item $\frac{p_m^{(1)}}{p_m}$ is the conditional probability of being infected given $m$.
	\item $p_m$ is the real proportion of people with the symptoms.
\end{itemize}
Let's assume without loss of generality, an ordering for the partition of $P$ by increasing severity and/or number of symptoms. Then we obtain the four following orderings:
\begin{align}
	&p(1) \leq \cdots \leq p(s) \leq \cdots \leq p(M), \label{inclusion}\\
	&p \left(1^{(1)} \right)  \leq \cdots \leq p \left(s^{(1)} \right)  \leq \cdots \leq p \left(M^{(1)} \right), \label{previnclusion}\\
	&\frac{p_1^{(1)}}{p_1} \leq \cdots \leq \frac{p_s^{(1)}}{p_s} \leq \cdots \leq \frac{p_M^{(1)}}{p_M}, \label{prevalence}\\
	&p_1 \geq \cdots \geq p_s \geq \cdots \geq p_M. \label{contra}
\end{align}

The intuition behind these is that the higher the degree and/or number of symptoms, then the higher the probability of being tested (\ref{inclusion}), the higher the probability of testing infected people (\ref{previnclusion}), the higher the probability of being infected inside that group (Eq. \ref{prevalence}), and the lower the real proportion of people with the symptoms (\ref{contra}).

From the conditional distribution of $(S^* | \textbf p^*, T = 1)$, we observe i.i.d.\ draws of $S | \textbf p$, whose density, because of Bayes theorem, is
\begin{align}
	f_{S | \textbf p}(s | \vec p) = f_{S^*| \textbf p^*, T}\left(s | \vec p, 1\right) 
								&=  \frac{P\left[ T=1 | S^* = s, \textbf p^* = \vec p \right]}{P\left[ T=1 | \textbf p^* = \vec p \right]} f_{S^*|\textbf p^*}\left( s |\vec p \right) \nonumber\\
	 							&= \frac{p (s)}{E[p(S^*) | \textbf p^* = \vec p]} p_s. \label{biasedS}
\end{align}

Assume there is no error in testing. Then we know exactly the proportion of infected people in the sample. Moreover, we know under which category $s$ is each person tested. Therefore, for all $s$, we can derive: 
\begin{align}
	f_{S| \textbf p} \left( s^{(1)}| \vec p \right) = f_{S^*| \textbf p^*, T}\left(s^{(1)}| \vec p, 1\right)
								&= \frac{P\left[ T=1 | S^* = s^{(1)}, \textbf p^* = \vec p \right]}{P\left[ T=1 | \textbf p^* = \vec p \right]} f_{S^*|\textbf p^*}\left( s^{(1)}|\vec p \right)\nonumber \\
								&= \frac{p \left(s^{(1)} \right)}{E[p(S^*) | \textbf p^* = \vec p]} p_s^{(1)}. \label{biasedSi} 
\end{align}

Thus, we obtained in (\ref{biasedS}) the biased estimate of the proportion of people tested ---and in (\ref{biasedSi}) the biased estimate of the proportion of prevalence) for each $s$---. The total biased estimator of tested people is
\begin{align}\label{totalbiasS}
	\sum_{s=1}^M  \frac{p (s)}{E[p(S^*) | \textbf p^* = \vec p]} p_s,
\end{align}
and the total biased estimator of prevalence is
\begin{align}\label{totalbiasSi}
	\sum_{s=1}^M \frac{p \left(s^{(1)} \right)}{E[p(S^*) | \textbf p^* = \vec p]} p_s^{(1)}.
\end{align}

\section{Bias correction}

The bias correction is easily determined.  It amounts to multiplying the quantity on the LHS of (\ref{biasedS}) and (\ref{biasedSi}) by the inverse of the quotient on the RHS of each respective equation. Specifically, the bias correction is given by $C(x) f_{S | \textbf p}(x | \vec p)$, where
\begin{align}\label{correction}
	C(x) \defeq \frac{P\left[ T=1 | \textbf p^* = \vec p \right]}{p(x)}.
\end{align}

Replacing $x$ by $s$ and $s^{(1)}$ will give us the bias correction for testing and prevalence, respectively, for each $s$. Summing over $s^{(1)}$ gives the sampling bias-corrected estimate of disease prevalence. Now, the numerator at the RHS of (\ref{correction}) can be estimated as $N_T/N$, where $N_T$ is the number of people tested, and $N$ is the census population. However, the denominator is unknown, but we can still say some things, depending on the the number of symptoms $M$ we are considering. 

\subsection{Big \textit s}\label{Bigs}

The bias problem of testing and prevalence is not in the last values of $S^*$, but in considering only the last values of $S^*$ for testing. This is the main thing to correct. We already learned that the overall proportion of people tested is $N_T/N$. We also know that most of the people tested are symptomatic; therefore, when $s$ approaches $M$, $\tilde p_s \defeq (N_T/N)  f_{S | \textbf p}(s | \vec p)$ is a good estimator of the real proportion of the last value. Finally, since we know that most of the symptomatic people will be infected, and most of them will get tested, then $\tilde p_s^{(1)} \approx \tilde p_s$ for large $s$.

\subsection{Small \textit s}\label{smalls}

When $s$ decreases to $1$, the situation is different. According to (\ref{inclusion}), in the absence of symptoms, the probability of being tested is small. Nonetheless, having few tests for small values of $s$ does not mean that the proportion of people infected/uninfected is close to 0. I.e., $p(s)$ might be small, but $p_s$ is large, which according to (\ref{biasedS}) is introducing a heavy bias.   Another way to think of this is that for an unbiased sample, $p_{s}$ would have a clear representation.  

Now, notice that, since $1 \geq C(s) f_{S | \textbf p}(s | \vec p)$,  it is possible to give a lower bound for $p(s)$:
\begin{align}\label{lowerbound}
	\frac{N_T}{N} f_{S | \textbf p}(s | \vec p) \leq p(s)
\end{align}

Giving $p(s)$ its lower possible value makes the corrected estimate $C(s)f_{S | \textbf p}(s | \vec p) = 1$. However, this implies that for all $s' \neq s$, the estimated probability of being tested is 0, which cannot be true. 

In order to solve this, let's divide $\mathcal P$ into three groups: $\mathcal P^- \cup \mathcal P^m \cup \mathcal P^+$, where $\mathcal P^-$ and $\mathcal P^+$ are the subsets of low and high values of symptoms, respectively; and $\mathcal P^m$, the symptoms in between.

According to Section \ref{Bigs}, we have already estimated the probability of the elements in $\mathcal P^+$. Thus making $p_+ \defeq \sum_{s \in \mathcal P^+} \tilde p_s$, we propose the following solution for values in $\mathcal P^-$:
\begin{enumerate}
	\item Take the space $\mathcal P^- \cup \mathcal P^m$, which has probability $1 - p_+$.
	\item Letting $\tilde M$ be the cardinality of the set $\mathcal P^- \cup \mathcal P^m$, assign equal probabilities $\tilde p_k$ to all $k \in \mathcal P^- \cup \mathcal P^m$. I.e., $\tilde p_s = \left(1- p_+ \right)/\tilde M$.
\end{enumerate}

The assignation of equal probabilities is justified by our total ignorance of the real proportions, resorting thus to Bernoulli's Principle of Insufficient Reason (PoIR, see, e.g., \cite{DembskiMarks2009a}), which says that in the absence of further knowledge we must assign equal probabilities to events. Or, more strongly, by the maximum entropy principle, since, according to Jaynes, ``in making inferences on the basis of partial information we must use that probability distribution which has maximum entropy subject to whatever is known. This is the only unbiased assignment we can make; to use any other would amount to arbitrary assumption of information which by hypothesis we do not have'' \cite{Jaynes1957a} (see also \cite{DiazMarks2020a}). In other words, if we are going to consider some other distribution than equiprobability, we must justify the reduction in entropy that is inserting bias.

\subsubsection{Prevalence}\label{prev}

We know that in the asymptomatic population, a non-negligible portion of individuals is infected. However, studies vary here as for the proportion of asymptomatic and infected individuals. Some of those studies maintain that most of the asymptomatic people are already infected. In such case, $p_s^{(1)} \approx p_s$, as we explained in Subsection \ref{smalls}. Others lean on the side that, for small, $s$, we have $p_s^{(1)}/p_s$ closer to 0 than to 1, but not necessarily approaching 0. We propose the following algorithm:
\begin{enumerate}
	\item If we don't have any information about the asymptomatic category with the disease, generate a random number $u$ according to a uniform distribution in the interval $(0,1)$.
	\item If we have some information about the asymptomatic category with the disease from the biased sample, make $u$ the proportion of prevalence inside the particular category of 		interest provided by the biased sample (i.e., $u = f_{S| \textbf p} \left( s^{(1)}| \vec p \right) /f_{S | \textbf p}(s | \vec p)$).
	\item Make $\tilde p_s^{(1)} = u\tilde p_s$.
\end{enumerate}

\begin{remark}
	We have randomized $u$ in $(0,1)$, assuming total ignorance regarding $p_s^{(1)}/p_s$. However, as we mentioned before the algorithm, other options are possible. For instance, in case of $p_s^{(1)}/p_s$ closer to 0 than to 1, but not necessarily approaching 0, it makes sense to randomize $u$ in $(0,1/2)$. 
\end{remark}

\begin{remark}
	It is also worth noticing that, because of step 2 in the previous algorithm, the importance of $u$ decreases as the information in the sample (particularly information about 			asymptomatic individuals and groups with lesser symptoms) increases.
\end{remark}

\subsection{Middle values of \textit s}

Notice that in the previous algorithm we are not asking to use $\tilde p$ for the correction in (\ref{correction}) of values in $S^m$ (although it might be done with some care.) This is because for these values we have few to no knowledge. Our recommendation is to collapse $\mathcal P$ to $\mathcal P^- \cup \mathcal P^+$. For this case, we explain the estimated values in Subsection \ref{M2}.

\subsection{Some model caveats}
The model is presented in some generality on purpose.  It naturally permits customization to different testing settings.  For instance, symptomatology may be extended to reflect different subpopulations (e.g., racial/ethnic subgroups, age groups, risk groups, environments). The model can also be generalized to index more than 1 type of test (not shown here).  

We also make the caveat that this is a {\it correction} of the bias, not a total elimination of it. This is a consequence of the fact that the correction factor $C(\cdot)$ in (\ref{correction}) cannot take any value, but has restrictions as explained in the obtention of equation (\ref{lowerbound}). Nonetheless, as we will see in the toy example and the two real-life scenarios in the last sections, the proposed method achieves very important reductions in sampling bias, particularly for the overall prevalence, but does not reduce it to zero.

\subsection{$M=2$}\label{M2}
As suggested before, an important simplification occurs when we consider only two groups of symptomatology: asymptomatic ($s=1$) and symptomatic ($s=2$). In this case, the analysis is conveniently reduced to: 
\begin{itemize}
	\item $\tilde p_2^{(1)} \approx \tilde p_2 = (N_T/N)  f_{S | \textbf p}(2 | \vec p)$ (from Subsection \ref{Bigs})
	\item $\tilde p_1 = 1 - \tilde p_2$ (from Subsection \ref{smalls}).
	\item $\tilde p_1^{(1)} = u\tilde p_1$ (from Subsubsection \ref{prev}).
\end{itemize}

\section{Estimated variance}

Let $\textbf X_1, \ldots, \textbf X_n$ be iid multinomial $\mathfrak M\left( 1; \textbf p^*\right)$. Then $\sum_{i=1}^n \textbf X_i \sim \mathfrak M( n; \textbf p^* )$. Let $\hat{\textbf p} \defeq n^{-1}\sum_{i=1}^n \textbf X_i$. By the MCLT, we know that
\begin{align}\label{normal}
	\sqrt n \cdot \hat{\textbf p} \sim AN\left(\textbf p^*, \Sigma/n\right),
\end{align}
where $AN$ stands for {\it asymptotically normal}, and $\Sigma =  \text{Diag}(\textbf p^*) - \textbf p^* {\textbf p^*}^T$. 

Now define 
\begin{align*}
	\textbf f_\textbf p \defeq \left( f_{S | \textbf p} \left( s_1^{(0)} | \vec p \right), f_{S | \textbf p} \left( s_1^{(1)} | \vec p \right), \ldots, f_{S | \textbf p}\left( s_M^{(0)} | \vec p \right), f_{S | 			\textbf p} \left(s_M^{(1)} | \vec p \right)\right),
\end{align*}	
and let $\textbf Y_1, \ldots, \textbf Y_n$ be iid $\mathfrak M\left( 1; \textbf f_\textbf p\right)$, so that $\sum_{i=1}^n \textbf Y_i \sim \mathfrak M( n; \textbf f_\textbf p)$. Applying the Delta method to $\hat{\textbf q} = n^{-1}\sum_{i=1}^n \textbf Y_i$, we obtain that $\sqrt n \cdot \hat{\textbf q} \sim AN(\textbf f_\textbf p, \textbf V_n)$, where $\textbf V_n = \frac{1}{n}g'(\textbf p) \boldsymbol \Sigma g'(\textbf p)^T$, and $g'(\textbf p)$ is
\begin{align}
	g'(\textbf p) = \frac{1}{P[T = 1 | \textbf p = \vec p]} \left( p \left(s_1^{(0)} \right), p \left(s_1^{(1)} \right), \ldots, p \left(s_M^{(0)} \right), p \left(s_M^{(1)} \right)  \right).
\end{align}

Notice that the correction in (\ref{correction}) applies $g^{-1}$ to $\textbf Y$, which leads back to $\textbf X$. Therefore, asymptotically, we obtain again the distribution in (\ref{normal}). 

In practice, we start with the biased iid $\tilde{\textbf Y}_1, \ldots, \tilde{\textbf Y}_{N_T}$, which are $\mathfrak M(1, \textbf f_\textbf p)$, whose sum is $\mathfrak M(n, \textbf f_\textbf p)$. Letting $\tilde{\textbf q}  = N_T^{-1} \sum \tilde{\textbf Y}_i$, we have again by the MCLT that $\sqrt{N_T} \cdot \tilde{\textbf q} \sim AN(\textbf f_\textbf p, \textbf V_{N_T})$. Making $h(\textbf f_\textbf p) = \textbf C\cdot \textbf f_\textbf p$, where $\textbf C$ is a vector with components as in (\ref{correction}), we can apply again the Delta method to $h$, obtaining
\begin{align}
	\sqrt{N_T}\cdot \tilde{\textbf q} \sim AN \left( h(\textbf f_\textbf p), \textbf C \textbf V_{N_T} \textbf C^T \right).
\end{align}

However, since we are not using $\textbf C$, but $\mathcal C$, an $M$-vector whose last component is $(N_T/N) f_M$ and with the first $M-1$ components being all $\frac{1}{M-1}\left( 1 - \frac{N_T}{N}f_M \right)$. Then, the variance-covariance matrix becomes 
\begin{equation*}
\begin{pmatrix}
	a\sigma^2_M & a\sigma^2_M & \cdots  & b\sigma^2_M \\
	a\sigma^2_M & a\sigma^2_M & \cdots  & b\sigma^2_M \\
	\vdots 	&	\vdots		& \ddots & \vdots \\
	a\sigma^2_M & a\sigma^2_M & \cdots  & b\sigma^2_M \\
	b\sigma^2_M & b\sigma^2_M & \cdots  & c\sigma^2_M 		
\end{pmatrix},
\end{equation*}
where $c= \left( \frac{N_T}{N} \right)^2$, $a = c\left( \frac{1}{M-1} \right)^2 $, $b = \frac{c}{M-1}$, and $\sigma^2_M$ is the variance of $f_M$. From this, the estimated variance of the total prevalence estimate can be calculated.  

\section{Toy Example}

Let's consider a population of size one million where individuals are labeled as asymptomatic ($S^* = 1$), few symptoms ($S^* = 2$), mild symptoms ($S^* = 3$), and all symptoms ($S^* = 4$). According to our previous definitions let's consider the following probabilities:
\begin{align*}
	p_1 &= 0.5				&	p_2 &=0.25				& 	p_3&=0.15				&	p_4&=0.1 \\
	p_1^{(0)} &=0.45			&	p_2^{(0)} &=0.2				&	p_3^{(0)} &=0.075			&	p_4^{(0)} &=0.01 \\
	p_1^{(1)} &=0.05			&	p_2^{(1)} &=0.05			&	p_3^{(1)} &=0.075			&	p_4^{(1)} &=0.09 \\
	p\left(1^{(0)} \right)&= 0.001	&	p\left(2^{(0)} \right)&= 0.01	&	p\left(3^{(0)} \right)&= 0.1		& 	p\left(4^{(0)} \right)&= 0.9 \\
	p\left(1^{(1)} \right)&= 0.001	&	p\left(2^{(1)} \right)&= 0.01	&	p\left(3^{(1)} \right)&= 0.1		& 	p\left(4^{(1)} \right)&= 0.9\\
	p(1) &= 0.001				&	p(2) &= 0.01				&	p(3) &= 0.1				&	p(4) &= 0.9
\end{align*}

None of these values is known to the researcher. The first row is the proportion of people in each group, it adds to 1. The second (third) are the population proportion of non infected (infected). Notice that the total prevalence is $\sum p_i^{(1)} = 0.265$.  The fourth is the proportion of tested people without the disease in each group. The fifth is the proportion of infected tested people in each group. The sixth is the resulting proportion of people tested within each group, it is derived using all the other rows. The proportion of tested people is
\begin{align}
	P[T = 1| \textbf p^* = \vec p] &= \sum_{i=1}^4 p(i) p_i = 0.108.
\end{align}

The researcher does not know this probability, but knows the total number of people tested (which can also be derived easily to be $N_T = 108,000$), and the overall census population $N=1,000,000$, with which she can estimate it. In addition, each individual's symptomatology status and test result is known.  Thus the proportion of people tested in each group is:
\begin{align*}
	f_{S| \textbf p} (1 |\vec p) &= f_{S^*| \textbf p^*, T}\left(1| \vec p, 1\right) \nonumber\\
				&=  \frac{P\left[ T=1 | S^* = 1, \textbf p = \vec p \right]}{P\left[ T=1 | \textbf p = \vec p \right]} f_{S^*|\textbf p}(1|\vec p)\nonumber \\
				&= \frac{0.001(0.5)}{0.108} \approx 0.004,\\
	f_{S| \textbf p} (2 |\vec p) &=\frac{0.01(0.25)}{0.108} \approx 0.023,\\
	f_{S| \textbf p} (3 |\vec p) &=\frac{0.1(0.15)}{0.108} \approx 0.14, \\
	f_{S| \textbf p} (4 |\vec p) &=\frac{0.9(0.1)}{0.108} \approx 0.833.
\end{align*}
Notice that the last term is grossly overestimated as $0.833$ (against the real 0.1), and the first two terms are grossly underestimated (against the real 0.5, 0.25); the third term is very well approximated, but in practice we do not know it. However, since we know that most of $N_T$ is made of the last group, $(N_T/N)0.833 = 0.0899$ becomes a very good estimator of its size, and $\tilde p_4^{(1)}$ is also 0.0899. According to our proposal for the small values of $S$, take $1 - (N_t/N)f_{S| \textbf p} (4 |\vec p) = 0.911$, and by  maxent, distribute it equally among the three remaining groups. In this way, we obtain a probability of 0.306 for each of the first three groups. According to Subsubsection \ref{prev}, using the information we have from the sample, we know that 50/500 were positive in the first group, 500/2500 were positive in the second group, and 7500/15000 were positive in the third group. In this way, $\tilde p_1^{(1)} = (50/500)0.306 = 0.0306$,  $\tilde p_2^{(1)} = (500/2500)0.306 = 0.0612$, and $\tilde p_3^{(1)} = (7500/15000)0.306 = 0.153$ are the corrected prevalence estimators for each of the remaining groups. From this, we obtain a total corrected prevalence of $0.0306 + 0.0612 + 0.153 + 0.0899 = 0.3347$. That is, we have obtained a big correction for the original estimator. The na\"ive estimator of prevalence produces $\sum_{i=1}^4 p\left(i^{(1)} \right)p_i^{(1)}/0.108 = 0.8245$.

This corrects greatly for $p_1$ and $p_2$, but is harmful for $p_3$, which is now overestimated, since its original na\"ive value was the right one, and the maxent value of 0.306 is doubling its real proportion. Moreover, since inside this third category the real number of infected is 75000, and in the fourth category it is 90000 (i.e., $7.5\%$ and $9\%$ of the overall population, respectively), any little change in the sampling scheme is likely to produce $\tilde p_3^{(1)} > \tilde p_4^{(1)}$. Since our correction averages on unknowns, this very situation is observed. Thus the toy example illustrates very well the method, even highlighting the comments in Subsection 3.3 about the difficulty of dealing with middle values. Nonetheless, the fact remains that the overall correction of prevalence is a very good one.

\section{The Diamond Princess COVID-19 outbreak: A real data example}\label{cruise}

A challenge in applying our bias correction to real data is that one does not know the population prevalence value typically (i.e. the true answer). However, we will consider the testing data from the Diamond Princess cruise \cite{MizumotoEtAl2020} in which essentially all people on the ship were tested, thus providing a ``population" prevalence value as the gold standard.   However, this example also has its problems in that it is considering mostly an elderly population. As background, a COVID-19 outbreak emerged on board the Diamond Princess cruise ship in early 2020. This was traced back to a former passenger who tested positive for the virus after disembarking in Hong Kong. After arriving in Yokohama, Japan, the ship was placed under quarantine and over a two-week period, essentially the entire population on the ship was tested by laboratory-based PCR for the virus. A total of 3063 unique tests were done out of 3711 total individuals. Some individuals were permitted to disembark at various points in time, but the status of those individuals not tested is assumed unknown.  

At the end of the testing period, out of 3603 tests conducted, there were 634 confirmed positive cases (positive for the virus).  Very importantly for our purpose, these were further categorized as 306 symptomatic and 328 asymptomatic at the time of testing.  We will assume that all PCR negative individuals were asymptomatic.  Additional demographic information is provided on country of residence, age and gender distributions for the cases \cite{MizumotoEtAl2020}.  The population value of prevalence found with PCR is thus 634/3063=0.206. We make the caveat that the accuracy of the PCR testing heavily depends on the delay between the time of infection and the time of sample collection, therefore, more probable than not, 0.206 does not correspond to the real prevalence. However, for the purposes of our illustration this is not an issue, and we will proceed considering it as the prevalence in the ship.

In order to demonstrate our bias correction,  we will analyze samples from the tested ship population considering four possible sampling protocols: 1) total bias in which we sample only the symptomatic positive individuals; 2) partial bias in which we draw a sample with $75\%$ representation from symptomatic individuals; 3) a balanced sampled of symptomatic and asymptomatic individuals; 4) a truly random sample from the ship's population based on population symptom frequencies.  

\vskip10pt

\noindent
{\bf Sampling Protocol 1}:  The sample consists only of the 306 symptomatic positive individuals.  The na\"ive estimate of prevalence is then 1.0, which is grossly overestimated.  The bias-corrected estimate of $\tilde{p}_{2}$ is $(N_{T}/N) 1 = 306/3063=0.099$.  This is also our bias-corrected estimate of $\tilde{p}_{2}^{(1)}$.  Furthermore, $\tilde{p}_{1} = 1-\tilde{p}_{2} = 1-0.099 = 0.901$ and, taking the average of $u$, we obtain $\tilde{p}_{1}^{(1)} = 0.5(0.901) = 0.450$.  Thus the bias-corrected total prevalence estimate is $\tilde{p}_{1}^{(1)} + \tilde{p}_{2}^{(1)} = 0.450 + 0.099 = 0.549$.

\vskip10pt
\noindent
{\bf Sampling Protocol 2}:  The sample consists of 306 symptomatic positive individuals and 101 asymptomatic ones.  Thus the sample has 75\% representation of symptomatic individuals.  We assume the number of positive asymptomatics in the sample is $101(328/3063) \approx 11$.  Thus the na\"ive sample estimate of prevalence is $(306+11)/407 = 0.779$.  Here $\tilde{p}_{2} = (N_{T}/N)(0.75) = 0.099$, which is also our estimate for $\tilde{p}_{2}^{(1)}$.   Then $\tilde{p}_{1} = 1 - \tilde{p}_{2} = 0.901$ and $\tilde{p}_{1}^{(1)} = (11/101)(0.901) = 0.098$.  Setting $u=11/101 \approx 0.109$ represents knowledge we have, which is the sample estimate of the prevalence for the asymptomatic group. Thus the biased corrected estimated of total prevalence is $\tilde{p}_{1}^{(1)} + \tilde{p}_{2}^{(1)} = .099 + .098 = 0.197$ which is very close to the true population value of 0.206.

\vskip10pt
\noindent
{\bf Sampling Protocol 3}:  The sample consists of 306 symptomatic positive individuals and 306 asymptomatic ones.   This means that all symptomatic individuals were still in the sample.  We will assume $306(328/3063) \approx 33$ positive cases among the asymptomatic individuals.  Thus the na\"ive estimate of prevalence is $(306+33)/612 = 0.554$.  Once again $\tilde{p}_{2} = (N_{T}/N)0.50 = 612/3063 * 0.50 \approx 0.099$, which is also our estimate for $\tilde{p}_{2}^{(1)}$.   Then $\tilde{p}_{1} = 1 - \tilde{p}_{2} = 0.901$ and $\tilde{p}_{2}^{(1)} = (33/306)(0.901) = 0.097$. Thus total corrected prevalence is 0.196, which is very close to the na\"ive estimate as we predicted.  

\vskip10pt

\noindent
{\bf Sampling Protocol 4}:  This is a random sample from the population.  Suppose we take $N_T = 500$.  Of these, $500(306/3063) \approx 50$ are symptomatic positive individuals and thus 450 are asymptomatic individuals.  Among the asymptomatic individuals, we assume $450(328/3063) \approx 48$ are positive for the virus.  Thus the na\"ive sample estimate of prevalence is $(50+48)/500 = 0.196$ which is quite close to the true value (save rounding errors).   Thus, we anticipate, the bias correction will do little.  Specifically, $\tilde{p}_{2} = (N_{T}/N)0.90 = 500/3063 * 0.90  \approx 0.147$ and this is also our estimate for $\tilde{p}_{2}^{(1)}$.  Then $\tilde{p}_{1} = 1 - \tilde{p}_{2} = 0.853$ and 
$\tilde{p}_{1}^{(1)} = 0.106(0.853) = 0.090$.  Thus the corrected prevalence estimate is $0.147+0.090 = 0.237$.  Again the true population value is 0.206.  

\vskip10pt
\noindent
{\bf Remark:}  One could also do brute force simulated bias estimation for each of the four sampling protocols above. This would mean drawing repeated samples of a given size from a ``ship population" characterized using particular protocol-specific probabilities for symptoms and then conditional on symptom status, population probabilities for being virus positive.   Corrected estimates of prevalence would be estimated for each sample and empirical biases estimated given that the true population prevalence is known.  The analyses conducted above can be thought of as based on an idealized representative sample for each protocol across all draws.  

\vskip10pt
\noindent
{\bf Remark:}  It would be of interest to do similar analyses broken out by age and gender.  However, the Diamond Princess data only provides aggregate information at these levels without a known mapping to symptomatology and viral presence status.  It's also important to note that all estimates and inferences are limited to considering the ship passengers as a population and should not be generalized further.

\section{Lombardy - Italy}

Since the share of symptomatic individuals on the Diamond Princess is controversial and the on-board population was highly skewed towards older ages, we add this new example on which the probability of developing symptoms is much lower (closer to the actual values estimated for COVID-19). This example will corroborate that the maxent principle (or Bernoulli's PoIR) works in a more representative scenario of the COVID-19 pandemic. The actual example comes from a recent preprint by Poletti et al, where the authors calculated the probability of symptoms and critical disease after SARS-CoV-2 infection in Lombardy, Italy \cite{PolettiEtAl2020}. 

In a sample of 5824 individuals it was possible to identify 932 infections through PCR testing. Moreover, besides these 932 infected individuals, they also detected 1892 infections using serological assays. Thus, the total of infected individuals was 2824. Among the total of infected, 876 were symptomatic ($31\%$). Therefore, since for our purposes we are only interested in detection by PCR and not through antibodies, we will not count the 1892 infections detected through serological assays. However, we will use the fact that $31\%$ of the cases were symptomatic assuming that the same percentage is holding for the 932 infections detected by PCR. Thus, in our case we will have a prevalence of $932/5824 = 0.16$; and among the infected, $0.31(932) = 289$ individuals will be symptomatic. The remaining 932-289 = 643 will be infected and asymptomatic.

\vskip10pt
\noindent
{\bf  Sampling Protocol 1}: The sample consists of the 289 infected and symptomatic individuals. In this case, the na\"ive estimate of prevalence is 1. The bias corrected estimate will be $\tilde p_2 = (N_T/N)1 = 289/5824 \approx 0.05$. And this will also be the correction for $\tilde p_2^{(1)}$. Then $\tilde p_1 = 1 - \tilde p_2 = 0.95$, and taking the mean of $u$ we obtain $\tilde p_1^{(1)} = 0.5(0.95) = 0.475$. Therefore, the total corrected prevalence is estimated as $\tilde p_1^{(1)} + \tilde p_2^{(1)} = 0.475 + 0.05 = 0.525$, which still high but corrects heavily the effects of a very bad sample.

\vskip10pt
\noindent
{\bf Sampling Protocol 2}: The sample consists of 384 individuals. Among these, 289 ($75\%$) are infected and symptomatic, and 95 ($25\%$) are asymptomatic. We assume the sample has $95(643/5824) \approx 10$ asymptomatic positive for the virus. So our na\"ive estimate of prevalence is $(289  + 10)/384 \approx 0.78$. In this case, $\tilde p_2^{(1)} = \tilde p_2 = (384/5824)0.75 \approx 0.049$. Now, $\tilde p_1 = 1 - \tilde p_2 = 0.951$; and setting $u$ as $10/95 \approx 0.105 $, $p_1^{(1)} = 0.105(0.951) \approx 0.1$. Therefore the total prevalence is corrected to $0.1 +0.049 = 0.149$, which is very close to the real 0.16.

\vskip10pt
\noindent
{\bf Sampling Protocol 3}: In this scenario we have 289 symptomatic positive and 289 asymptomatic. We assume the sample has $289(643/5824) \approx 32$ asymptomatic and infected individuals. The na\"ive estimate is $(289 + 32)/578 \approx 0.55$. However, $\tilde p_2^{(1)} = \tilde p_2 = (578/5824)0.5 \approx 0.05$; and $\tilde p_1 = 0.95$. In this case, $\tilde p_1^{(1)} = (32/289)0.95 \approx 0.105$., where $u = 32/289$. Therefore, $0.105 + 0.05 = 0.11$ is the corrected estimate of prevalence.

\vskip10pt
\noindent
{\bf Sampling Protocol 4}: This sample is truly random. Say $N_T = 600$. Among these, $600(289/5824) \approx 30$ are symptomatic and positive. Therefore, 570 are asymptomatic. Among the asymptomatic group, we are going to assume $600(643/5824) \approx 66$ infected individuals. The na\"ive sample estimate is thus $(66+30)/600 = 0.16$, which of course is the same as the real prevalence. In this case, the correction will work like this: $p_2^{(1)} = p_2 = (600/5824)0.95 \approx 0.098$. Then $\tilde p_1 = 1- \tilde p_2 = 0.902$, and $p_1^{(1)} = (66/570)(0.902) \approx 0.1044$. Therefore, the total prevalence is estimated as $0.1044 + 0.098 = 0.2024$. In this case, the correction is not bad, but of course does not do as well as the truly random sample.

\section{The need for further bias correcting and discussion}
The model assumes tests with no errors (i.e. false positives or false negatives).  Clearly this is not the situation in practice.  Specificities and sensitivities can often be less than ideal.  Sampling bias-corrected estimates of prevalence can be further corrected in a second stage using the methods of  \cite{Diggle2011} or \cite{Greenland1996} which account for using imperfect tests.   

Our study demonstrates that under biased sampling designs that are often difficult to avoid  in testing studies for COVID-19, the resulting biased estimates of prevalence can be corrected using simple methodology derived and adapted from corrections for publication bias used in meta analysis studies.  Further research is needed in order to correct the prevalence of ``middle'' groups, as stated in Subsection 3.3 and seen with the overestimation of $s=3$ in the toy example. However, the correction detailed in our toy example, while extreme, shows the effectiveness of the correction for the total prevalence.  The corrections can be used directly in practice using the data collected even though many of underlying quantities in the population may be unknown to the researcher.

\section{Acknowledgments}
JSR was partially supported by NSF grant DMS-1915976 and NIH grants U54 MD010722 and UL1 TR000460. We would like to thank the two referees for their helpful input which improved the final version of the paper.

\bibliographystyle{plainnat}
\bibliography{TestingBias}

\end{document}